\documentclass[prc,aps,preprint]{revtex4-1}
\usepackage{graphicx}
\usepackage{multirow}
\usepackage{natbib}
\begin{document}

\title{Projection of angular momentum via linear algebra}

\author{Calvin W. Johnson}
\affiliation{Department of Physics, San Diego State University,
5500 Campanile Drive, San Diego, CA 92182-1233}
\author{Kevin D. O'Mara}
\affiliation{Department of Physics, San Diego State University,
5500 Campanile Drive, San Diego, CA 92182-1233}

\begin{abstract}
Projection of many-body states with good angular momentum from an initial state is usually accomplished by a 
 three-dimensional integral.  
We show how projection can instead be done by solving a straightforward system of 
linear equations. We demonstrate the 
method and  give sample applications to   $^{48}$Cr and  $^{60}$Fe in the $pf$ shell. 
This new projection scheme, which is competitive against the standard numerical quadrature, should also be  applicable to other quantum numbers such as 
isospin and particle number.
\end{abstract}
\maketitle

\section{Introduction}

In the quantum theory of many-body systems we can have  exact or nearly exact symmetries, sometimes referred to as `good' symmetries, such as angular momentum, parity, 
isospin (for nuclear physics), and particle number.     Breaking those 
good symmetries can paradoxically improve results and insights. In mean-field methods such as the Hartree-Fock (HF) approximation, deformed solutions often have lower energies than 
spherically symmetric solutions; number-mixing appproximations such as Hartree-Fock-Bogoliubov and  Bardeen-Cooper-Schriefer also 
improve estimates of the ground state \cite{ring2004nuclear}.  

To go even further, one wants to restore or project good quantum numbers, either before or after variation.  
 (Henceforth we will restrict our discussion to angular momentum, though certainly our approach could be applied to 
other quantum numbers.) 
 Angular momentum projection is typically accomplished by a three-dimension integral over the Euler angles \cite{ring2004nuclear}, using  a complete, orthogonal set of 
angular functions.  

 We present an alternate approach which, instead of relying upon orthogonality, only needs two conditions: first, that the 
eigenfunctions of rotation are merely linearly independent, and second, that any initial state contains
only a finite number of angular momenta.  Under these two conditions  projection of angular momentum 
in a finite space can be cast as solving a straightforward set of linear equations. The first condition is already trivially satisfied from orthogonality of the rotation 
matrices;  and
in a finite space  not only can there be  only  a finite set of angular momentum eigenfunctions,  in fact in 
most cases Slater determinants calculated from mean-field theory contain only a small fraction of the possible angular momentum states.
Because of this small number we find  linear algebra projection of angular momentum can be numerically competitive with the standard three-dimensional integral.
Finally, we also discuss how the numerical efficiency can be improved by using the norm or overlap matrix elements to reduce the dimension of the 
linear algebra to be solved.

First we review the standard projection via quadrature over orthogonal functions. 
The most common projection operator uses the fact that rotation of a state of good total angular momentum $J$ and $z$-component $K$, 
(in nuclear physics, one often uses $K$ to denote the $z$-component of angular momentum in the so-called ``intrinsic'' frame \cite{ring2004nuclear}; here we use it to 
denote the $z$-component in the original frame, in order to better match common representations in the literature),  does not mix total angular momentum $J$ but does mix 
the $z$ component \cite{edmonds1996angular}:
\begin{equation}
\hat{R}(\alpha, \beta,\gamma) | J \, K \rangle = \sum_{M} {\cal D}^{(J)}_{M,K} (\alpha, \beta,\gamma)| J \,M \rangle,
\end{equation}
where we have the the rotation operator over the Euler angles
\begin{equation}
\hat{R}(\alpha, \beta,\gamma) = \exp \left( i \gamma \hat{J}_z \right ) \exp \left( i \beta \hat{J}_y \right )  \exp \left( i \alpha \hat{J}_z \right ),
\end{equation}
with $\hat{J}_z$ and $\hat{J}_y$  the generators of rotations about the $z$ and $y$-axes, respectively, and 
where ${\cal D}^{(J)}_{M,K}$ is a \textit{Wigner D-matrix}.  
The Wigner $D$-matrices are the matrix elements of the the rotation operator in a basis of good angular momentum, and 
 can be shown to be eigenfunctions 
of the quantized symmetric top \cite{edmonds1996angular} and 
form a complete orthogonal set,
\begin{equation}
\int_0^{2\pi} d\alpha  \int_0^\pi \sin \beta d\beta   \int_0^{2\pi} d\gamma  \,
{\cal D}^{(J^\prime)*}_{M^\prime, K^\prime} (\alpha, \beta,\gamma) {\cal D}^{(J)}_{M,K} (\alpha, \beta,\gamma)
= \frac{8\pi^2}{2J+1} \delta_{J,J^\prime}\delta_{M,M^\prime}\delta_{K,K^\prime},
\label{Dorthog}
\end{equation}
a property 
which can be exploited for angular momentum projection.  Consider some initial state which is a mixture of states of good angular momenta:
\begin{equation}
| \Psi \rangle = \sum_{J,\lambda} c_{J,\lambda} | \psi: J \lambda, K_\lambda \rangle.  \label{JKexpansion}
\end{equation}
We use $\lambda$ to distinguish components with the same $J$ but different initial $K_\lambda$; in general these will not be eigenstates 
of the Hamiltonian.
 Then applying the rotation operator,
\begin{equation}
\hat{R}(\alpha, \beta,\gamma) | \Psi \rangle = \sum_{J,\lambda} c_{J,\lambda} \sum_M {\cal D}^{(J)}_{M,K_\lambda} (\alpha, \beta,\gamma) | \psi: J \lambda, M \rangle,
\label{applyrotation}
\end{equation}
where we use $K_\lambda $ to denote the value of $J_z$ in the original state, and $M$ in rotated states.
The reason we do this is in the expansion (\ref{JKexpansion}), if we rotate states with the same $J$ and different $\lambda$ to have the same orientation, they 
need not be orthogonal to each other, that is, $\langle  \psi: J \lambda , M  | \psi: J \mu, M \rangle$ does not need to vanish when 
$\lambda\neq \mu$.  
To project angular momentum, we rotate all components of (\ref{JKexpansion}) to have the same orientation $M$. 
 This leads to the standard angular 
momentum projection equations \cite{ring2004nuclear}:  one constructs the norm matrix
\begin{equation}
N^J_{MK} =   \frac{8 \pi^2}{2J+1} \int d\Omega \, {\cal D}^{(J)*}_{M,K}(\Omega) \left \langle \Psi \left | \hat{R}(\Omega) \right | \Psi \right \rangle \label{normdefnquad} 
\end{equation}
where $\Omega$ stands in for the Euler angles $\alpha, \beta, \gamma$. The norm matrix can be 
 can be written in terms of the expansion (\ref{JKexpansion}):
\begin{equation}
N^J_{MK} = \sum_{\lambda \mu} \delta_{M, K_{\lambda} } c^*_{J,\lambda} c_{J,\mu}  \langle \psi: J \lambda, M | \psi: J \mu, M \rangle.\label{normdefn} 
\end{equation}
We can do the same for the Hamiltonian matrix
\begin{eqnarray}
H^J_{MK} =   \frac{8 \pi^2}{2J+1} \int d\Omega \, {\cal D}^{(J)*}_{M,K}(\Omega) \left \langle \Psi \left | \hat{H} \hat{R}(\Omega) \right | \Psi \right \rangle \label{hamdefn} \\
= \sum_{\lambda \mu}  \delta_{M, K_{\lambda} }  c^*_{J,\lambda} c_{J,\mu}  \langle \psi: J \lambda, M | \hat{H} | \psi: J \mu, M \rangle, \nonumber
\end{eqnarray}
where $\hat{H}$ is the many-body Hamiltonian. One then solves for each $J$ the generalized eigenvalue problem, with solutions labeled by $r$
\begin{equation}
\sum_K H^J_{MK}g^{(J)}_{K,r} = E_r \sum_K N^J_{M,K}g^{(J)}_{K,r}, \label{eigen}
\end{equation}
with the reconstructed eigenfunction
\begin{eqnarray}
|\Psi J M,r \rangle = \sum_K g^{(J)}_{K,r} \int d\Omega \, {\cal D}^{(J)*}_{M,K}(\Omega) | \Psi \rangle \\
=  \sum_{K} g^{(J)}_{K,r}  \sum_\lambda \delta_{K, K_\lambda} c_{J, \lambda} | \psi: J \lambda,M \rangle.
\nonumber
\end{eqnarray}
If one projects from a single initial state, for each $J$ there are at most $2J+1$ unique solutions; the number of 
actual unique solutions corresponds to the number of nonzero eigenvalues of the matrix $\mathbf{N}^{J}$, although 
in many applications one projects on multiple initial states.
%In practice the eigenvalues of the norm matrix do not go sharply to zero; one of the questions that concerned us was the robustness of our new method if one excluded 
 %small but still nonzero norm eigenvalues. 

Some of the many applications are projected Hartree-Fock \cite{ring2004nuclear,PhysRev.156.1087} including variation after projection \cite{ring2004nuclear,PhysRevC.71.044313},
 and Hartree-Fock-Bogoliubov \cite{ring2004nuclear,sheikh2000symmetry,Borrajo2016};
the Monte Carlo Shell Model \cite{PhysRevLett.77.3315,abe2013recent}; the projected shell model \cite{hara1995projected,sun1997fortran,PhysRevLett.82.3968}; the
projected configuration-interaction \cite{PhysRevC.79.014311} and related methods \cite{schmid2004use}; and 
projected generator coordinate \cite{PhysRevC.65.024304,PhysRevC.81.044311}. 
This list is far from exhaustive.  

The matrices $N^J_{MK}$ and $H^J_{MK}$  are generally small in dimension, but to arrive at them one needs to evaluate
 the integrand matrix elements $\left \langle \Psi \left | \hat{R}(\Omega) \right | \Psi \right \rangle$ and $\left \langle \Psi \left | \hat{H} \hat{R}(\Omega) \right | \Psi \right \rangle$ for a large number of angles $\Omega$.
%Using Gauss-Laguerre quadrature, if one uses unrestricted starting wavefunctions and if one wants to extract high $J$ typically one needs around 25 points for each Euler angle,  or more than 15,000 mesh points.  
As an example, a recent paper \cite{Borrajo2016} used 32 points per Euler angle, or a total of $32^3=32,768$ angles.
If one imposes symmetries, i.e. axial symmetry,  upon the mean-field state one can reduce the number of evaluations \cite{hara1995projected}, but even 
so  each evaluation is computationally intensive, especially of the Hamiltonian; see section \ref{need2know}. 

Projecting out from fully triaxial states, or projecting additional quantum numbers such as isospin or particle number, is so computationally intensive
one often has to  severely restrict the model space \cite{PhysRevC.92.064310}.  Given the applications of angular momentum projection and the computational burden, we were motivated to find an alternate approach, not by speeding up 
the evaluation of the integrands for any set of Euler angles, but rather to reduce the number of mesh points needed.  
%In the following section we describe our new method.

\section{Linear algebra solution for angular momentum projection}

Equations (\ref{normdefnquad}) and (\ref{hamdefn}) are usually taken as recipes for computing the norm and Hamiltonian matrices, respectively. We ignore the integrals,
instead taking (\ref{normdefn}) and (\ref{hamdefn})
 them as definitions of those matrices in terms of the expansion (\ref{JKexpansion}).  Starting from Eqn.~(\ref{applyrotation}) and using those definitions, one finds 
for any given value of the Euler angles $\Omega = (\alpha, \beta, \gamma)$,
\begin{equation}
\langle \Psi | \hat{R}(\Omega) | \Psi \rangle = \sum_{J,K,M}  {\cal D}^{(J)}_{M,K} (\Omega)N^J_{MK}, \label{linrelnorm} 
\end{equation}
\begin{equation}
\langle \Psi | \hat{H} \hat{R}(\Omega) | \Psi \rangle = \sum_{J,K,M}  {\cal D}^{(J)}_{M,K} (\Omega) H^J_{MK}. \label{linrelham} 
\end{equation}
These   key equations say
 $\langle \Psi | \hat{R}(\Omega) | \Psi \rangle$ is a linear combination of the the norm matrix elements $N^J_{MK}$, and 
the same for  $ \Psi | \hat{H} \hat{R}(\Omega) | \Psi \rangle $ and the Hamiltonian matrix elements $H^J_{MK}$.  While in usual practice 
one uses the orthogonality of the ${\cal D}$-matrices, Eq.~ (\ref{Dorthog}), 
to find $N^J_{MK}$, $H^J_{MK}$, we instead rely only upon their linear independence and solve 
 solve Eqn.~(\ref{linrelnorm}) and (\ref{linrelham}) as a linear algebra problem. 
That is, if we label a particular choice of Euler angles $\Omega$ by $i$ and the angular momentum quantum numbers $J,M,K$ by $a$, 
and define 
\begin{eqnarray}
n_i \equiv \langle \Psi | \hat{R}(\Omega_i) | \Psi \rangle , \nonumber\\
D_{ia} \equiv  {\cal D}^{(J_a)}_{M_a,K_a} (\Omega_i), \\
N_a \equiv N^{J_a}_{M_a K_a}, \nonumber 
\end{eqnarray}
we can rewrite Eq.~(\ref{linrelnorm}) simply as 
\begin{equation}
n_i = \sum_a D_{ia} N_a \label{compactLA}
\end{equation}
which can be easily solved for $N_a = N^J_{M,K}$, as long as $D_{ia}$ is invertible, with a similar rewriting of Eq.~(\ref{linrelham}) and 
solution for $H^J_{M,K}$.

A key idea is that the sums (\ref{linrelnorm}), (\ref{linrelham}) are finite.   To justify this, we introduce 
the  fractional `occupation' of the wave function with angular momentum 
$J$, which is the trace of the  fixed-$J$ norm matrix:
\begin{equation}
f_J = \sum_{M} N^J_{M,M}. \label{def_fJ}
\end{equation}
Assuming the original state is normalized, one trivially has 
\begin{equation}
\sum_J f_J = 1.
\label{sumrule}
\end{equation}  

The fractional occupation $f_J$ and its sum rule (\ref{sumrule}) have multiple uses. First, the sum rule is an important check on any calculation. 
 Second and more important, one can use the exhaustion of the sum rule to determine a maximum angular 
momentum, $J_\mathrm{max}$, 
in our expansions; in our trials we found both (\ref{JKexpansion}) and  (\ref{sumrule})  dominated by a finite and relatively small number of terms,
 far fewer terms than are allowed even in finite model spaces.  As discussed in the next section, we found that fractional occupations below 
 $0.001$ could be safely ignored.

\begin{figure}
\centering
\includegraphics[scale=0.45,clip]{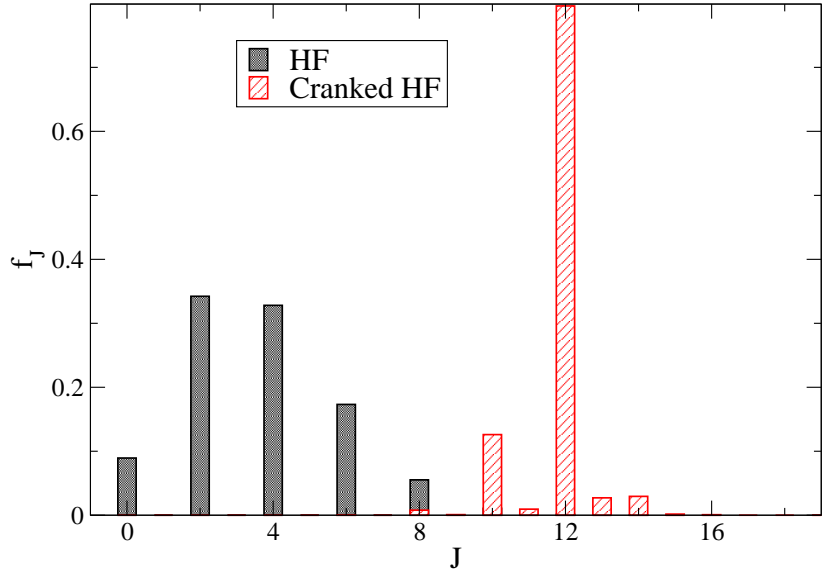}
\caption{(Color online) 
Fraction of given angular momentum $J$ ($f_J$) in a Slater determinant, calculated both for ordinary Hartree-Fock (HF) and cranked HF with 
cranking frequency $\hbar \omega = 1.0$ MeV, for $^{60}$Fe in the $pf$ shell. The maximum angular momentum for this nuclide in this model space is 
26 $\hbar$.}
\label{Jdistro}
\end{figure}

In general, for a Hartree-Fock state the distribution of $f_J$ is weighted towards low $J$ and does not reach the maximum $J$ in the many-body space. 
In Fig.~\ref{Jdistro} we show this for $^{60}$Fe in the $pf$ shell where the maximum $J$ in the space is $26$, using the $pf$-shell interaction  derived from 
a $G$-matrix, version A, or GXPF1A \cite{PhysRevC.65.061301,honma2005shell}. 
 We also show the distribution of $f_J$ for a strongly cranked 
Slater determinant, where we added $\hbar \omega J_z$ (or, alternately, $J_x$) with $\hbar \omega = 1.0$ MeV.  For uncranked Hartree-Fock the maximum $J$ is 
12 with an $f_J \sim 0.001$; because the Hartree-Fock state is axially symmetric, only even $J$ were populated. For the strongly cranked Hartree-Fock state, most of the state had $J=12$, but 
the range was between 6 and 16. Because the cranked HF state was triaxial we also got odd $J$.   

%\textit{Need to put elsewhere:} 
%We have implemented this idea and confirmed that it works, gives robust results agreeing with those from quadrature, and reduces the number of evaluations 
%of the left hand sides of Eqn.~(\ref{linrelnorm}) and  (\ref{linrelham}) by up to an order of magnitude or more, leading to a proportional speed up in projection. 

Our method is not completely unprecedented. For example, previous applications in the so-called shell-model Monte Carlo  extracted traces over states with 
good  particle number  \cite{PhysRevC.49.1422} and good $M$ ($z$-component 
of angular momentum) \cite{PhysRevLett.99.162504} via Fourier methods which can be thought of as inverting the linear relation analytically. To the best of our knowledge, however,
this is the first time one has fully projected out angular momentum using inversion of linear equations.

%Furthermore, when $| \Psi \rangle$ is a solution to some variational method, the amplitudes $c_{JK}$ are usually peaked for small values of $J$, which means the number of non-vanishing amplitudes is fewer than the maximum possible in the finite space. 

\section{Implementation}

To implement projection by linear algebra, we worked in finite single-particle shell-model spaces, such as the $pf$ shell. 
We used the code {\tt SHERPA} \cite{PhysRevC.66.034301,stetcu2003toward} to generate unrestricted Hartree-Fock 
states $| \Psi_{HF} \rangle$;  {\tt SHERPA} reads in shell-model configuration-interaction compatible files, and can handle even and odd number of protons 
and neutrons, and allows for arbitrary triaxility.  We then projected out the norm and Hamiltonian matrices using both quadrature and linear inversion.

%In both methods of projection, the computational burden depends upon the number of evaluations required.  

There are two practical choices which must be made. The first is one of tolerance of small values; the second is the choice of mesh of 
Euler angles for evaluating matrix elements. 

When solving the generalized eigenvalue equation (\ref{eigen}), the norm matrix often is 
not formally invertible, because it has eigenvalues which either are zero or are very small.   Such tiny eigenvalues generally have numerical noise 
and including them leads to unphysical solutions.  Hence one needs to choose a tolerance $\epsilon$; for any eigenvalues less than $\epsilon$ 
we exclude the associated subspace. We found a tolerance of $\epsilon\sim 0.001$ worked satisfactorily.   One also has to choose a 
tolerance for satisfying the sum rule (\ref{sumrule}). This essentially dictates the maximal $J$ used in inversions. Again, we found that a
tolerance of $\sim 0.001$ worked satisfactorily, that is,  $\sum_{J}^{J_\mathrm{max}}f_J \geq 0.999$ determined $J_\mathrm{max}$.

In order to solve for the projected Hamiltonian and norm matrices, one must first choose a mesh of Euler angles such that the linear equations are solvable. 
For our initial inversions, we found a simple mesh, which allowed us to invert each Euler angle separably,  worked well. 
To simplify our solution, we solved for each each quantum number $J, K, M$ separately.  To make clear this approach, we 
we expand (\ref{compactLA}) to read
\begin{equation}
n_{ijk} = \sum_{JKM} D_{ijk,JKM} N^J_{KM} \label{expandedLA}
\end{equation}
where
\begin{eqnarray}
n_{ijk} \equiv  \langle \Psi |  \exp( i \alpha_i \hat{J}_z) \exp( i \beta_j \hat{J}_y) \exp( i \gamma_k \hat{J}_z)  | \Psi \rangle \nonumber \\
= \sum_{JKM} D^{J}_{K, M} ( \alpha_i, \beta_j, \gamma_k) \, N^{J}_{K, M} \nonumber \\
= \sum_{JKM} e^{i\alpha_i M} d^J_{MK}(\beta) e^{i \gamma_k K}\,N^{J}_{K, M}
\end{eqnarray}
where $d^J_{MK}(\beta)$ is of course the Wigner little-$d$ function. 
By separable  we mean the Euler angles $\alpha_, \beta_j, \gamma_k$ run independent of each other and we solve (\ref{expandedLA}) one index at a time. 
To begin with, we use on the angles $\alpha, \gamma$ an equally spaced mesh
$\gamma_k = (k-1)\frac{2 \pi }{2J_\mathrm{max} + 1}, k = 1, 2J_\mathrm{max} + 1$, (and similarly for $\alpha_i$).
We  analytically invert the 
finite Fourier sums using  \cite{PhysRevLett.99.162504}
\begin{equation}
\frac{1}{N} \sum_{k=1}^N \exp \left ( i \frac{2\pi M k}{N}  \right ) = \delta_{M,0}.
\end{equation}
and introduce the matrix 
\begin{equation}
Z_{Kk} = \frac{1}{2J_\mathrm{max} + 1} \exp( -i K \gamma_k ). \label{inverseM}
\end{equation}
to arrive at the 
intermediate  quantity
\begin{equation}
n_{j,MK} = \sum_{ik} Z_{Mi} Z_{Kk} N_{ijk} = \sum_J d^J_{MK}(\beta_j) N^J_{MK}. \label{jbetatransform}
\end{equation}
At this point we have done two-thirds of the work. We have `projected' the magnetic quantum numbers $K$ and $M$; all that remains is to project out total $J$.

To obtain projection on total $J$, we also need a mesh on $\beta$. We chose  an equally spaced mesh on $\beta_j$:
\begin{equation}
\beta_j = (j-1) \frac{\pi}{N}, \label{betamesh}
\end{equation}
where $N= J_\mathrm{max}+1$ if an even system and $=J_\mathrm{max}+1/2$ if an odd number of nucleons.
To invert, construct 
\begin{equation}
\Delta^{J^\prime J}_{MK} = \sum_j d^{J^\prime}_{MK}(\beta_j) d^{J}_{MK}(\beta_j), \label{DeltaJJdefn}
\end{equation}
with $J, J^\prime \geq |M|, |K|$. 
(This step is inspired by singular value decomposition treatment of nonsquare matrices, although we are not formally carrying out singular value decomposition.)
The matrix $\Delta^{J^\prime J}$ is real and symmetric with fixed $M, K$.  We numerically confirmed, for $J \geq \max (|K|, |M|)$, it is invertible and has nonzero (and nonnegative) eigenvalues. 
We construct another intermediate matrix, 
\begin{equation}
\tilde{N}_{J^\prime, MK} = \sum_j d^{J^\prime}_{MK}(\beta_j) N_{j,MK}. \label{adjointsvd}
\end{equation}
Then we simply solve 
\begin{equation}
\sum_J   \Delta^{J^\prime J}_{MK}   N^J_{MK} = \tilde{N}_{J^\prime, MK}  \label{finaleqn}.
\end{equation}
In a similar fashion we solved for $H^J_{KM}$ and then could solve Eq.~(\ref{eigen}).  We confirmed our matrices were the same using either quadrature of 
linear algebra to project.

\section{Example applications}

\label{examples}

\begin{table}
\caption{Low-lying energies, in MeV, of $^{48}$Cr from projected Hartree-Fock in the $pf$ shell with the interaction GX1A. 
$f_J$ is the fraction of the Hartree-Fock state with angular momentum $J$, eqn.~(\ref{def_fJ}).  `quad' refers to projection by 
 quadrature, eqn.~(\ref{normdefnquad}) and (\ref{hamdefn}) and the number of points, while `LAP' refers to linear algebra projection.
\label{cr48}}
\begin{tabular}{|r|r|r|r|c|c|}
\hline
$\,\,  J\,\, $  & $f_J$&  quad.  & quad.  & LAP  & LAP  \\
      &  &  20 pts &  40 pts &  (full)      &  (`need-to-know' )   \\
\hline
 0 &  0.0695 & -97.9760 & -97.9778 &    -97.9778   & -97.9775  \\
 \hline
 2  &   0.2817 & -97.5024 &   -97.5044  &  -97.5044   & -97.5037  \\
\hline
 4   &  0.3115 & -96.5570 &  -96.5522  &    -96.5522   & -96.5520   \\
\hline
 6   &  0.2077  & -95.2162 & -95.1115 &   -95.1115  &   -95.1114   \\
\hline
 8   & 0.0935   & -104.2016 &  -93.3330 &  -93.3330   & -93.3325  \\
\hline
10   &  0.0292 & -133.9126 &  -91.1718    & -91.1719   & -91.1717 \\
\hline
12    &   0.0069  & -141.5208 &  -88.8433         &  -88.8499   & -88.8558 \\
\hline
\end{tabular}
\end{table}

We give two brief example applications of the method, both in the $pf$ shell. The first, $^{48}$Cr, shows the level of agreement between projection by quadrature and 
linear algebra projection. We have done numerous other tests in other nuclei and other model spaces, including multi-shell spaces, and find similar agreement.
 The second exhibits good agreement of yrast excitation energies in  $^{60}$Fe between full shell-model diagonalization
and linear algebra projected Hartree-Fock. 

Table \ref{cr48} shows the low-lying projected Hartree-Fock spectra of $^{48}$Cr computed in the $pf$ shell
with  GXPF1A \cite{PhysRevC.65.061301,honma2005shell}. 
We show four calculations, two quadrature calculations with 20 and 40 points, and two linear algebra projection (LAP) calculations, one with a full inversion 
and one with the `need-to-know' modification described in section \ref{need2know}.  The quadrature calculation with 40 points on each angle, 
or $40^3 = 64,000$ evaluations agrees with both the full LAP (12,615 evaluations) and the need-to-know (4,375 evaluations) to within 2 keV, 
except for the $J=12$ state, which agreed within $\sim 10$ keV.   (This latter had a fractional occupation $f_J = 0.007$, close to our 
tolerance, and in general we found less reliable results in both quadrature and LAP for states with tiny fractional occupations.) 
The quadrature calculation with 20 points on each angle, or 8000 evaluations, agrees for small $J$ but breaks down for larger angular momentum;
although we don't show it, this breakdown is also signaled by a failure in the sum rule (\ref{sumrule}).

\begin{figure}
\centering
\includegraphics[scale=0.45,clip]{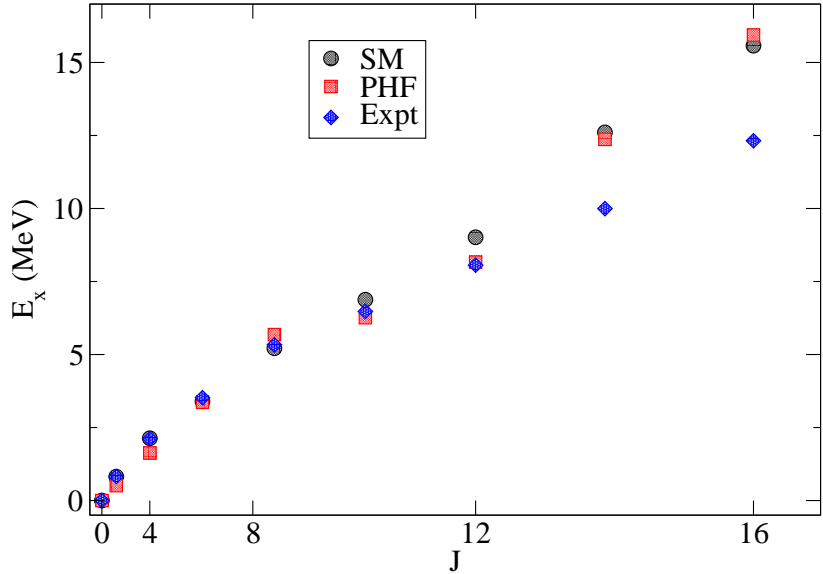}
\caption{(Color online) Yrast excitation energies for $^{60}$Fe computed in the $pf$ shell with the semi-phenomenological interaction GXPF1A. 
Results are from full shell-model diagonalization (SM, black circles) and linear algebra projected Hartree-Fock (PHF, red squares). For comparison
we also show the experimental values (blue diamonds).
.}
\label{yrast}
\end{figure}

The other demonstration is of the yrast excitation energies of $^{60}$Fe, shown in Fig.~\ref{yrast}, also computed in the $pf$ shell with the GXPF1A interaction.
We compare results from full shell-model diagonalization (black circles), also known as configuration-interaction method, 
against our LAP projected Hartree-Fock results (red squares).
For such a simple calculation we get  good agreement between the two calculations, although both diverge at high $J$ from experiment (blue diamonds).

Although we have shown only even-even cases, LAP works just fine for odd-$A$ and odd-odd cases. 
Of course, projected Hartree-Fock spectra do not always provide a good approximation to equivalent full shell-model diagonalization.  Unsurprisingly 
the best agreement was for rotational spectra of even-even nuclei. We plan to study this more systematically in the future. 
In general the PHF spectra for even-even 
nuclei better approximate the numerically exact results; in addition, systems with an odd number of particles generally mix in all values of 
$J$ and $M$, thus making the need-to-know algorithm less applicable. Nonetheless, we have found excellent agreement between quadrature 
projection and LAP for odd-$A$ and odd-odd nuclei.

We carried out similar explorations for many nuclides in the $sd$- and $pf$ shells, in the $sd$-$pf$ space, in the $0g_{7/2}$-$1d$-$2s$-$0g_{11/2}$ space, and 
in a no-core shell model space including all orbits up to principal quantum number $N=5$.  This included odd-$A$ and odd-odd nuclides.  Qualitatively all results 
were similar to Fig.~\ref{Jdistro}, and without cranking we seldom found $f_J > 0.001$ for $J > 16$. (In spaces including opposite parity orbits we also projected on parity, by taking 
$| \Psi  \rangle \pm P | \Psi \rangle$ where $P$ is the parity inversion operator.) Only when we cranked  did we get large $J$, but in those cases the distribution was again clustered on a relatively small number of $J$ values.

\section{Computational burden and improved efficiency through `need to know'}

\label{need2know}

The  motivation for introducing this new algorithm is to reduce the computational burder of projecting good quantum numbers. In this section 
we discuss the origin and scaling of the computational burden and outline an advanced algorithm with even greater 
efficiency.

Let's briefly overview some details on computing matrix elements in this particular framework \cite{PhysRevC.48.1518}.
 Suppose, starting from $N_s$ orthonormal single-particle basis states $\phi_a$, $a=1,N_s$, represented by 
creation and destruction operators $\hat{c}^\dagger_a, \hat{c}_a$, we construct general single-particle states
$\sum_a \Psi_{ai} \phi_a$, $i=1,N_p$. Then we can represent the Slater determinant which is the antisymmetrized product of these $N_p$ states by the
 $N_s \times N_p$ matrix $\mathbf{\Psi}$, even if the column vectors are not orthonormal.  The overlap between two such general Slater determinants is 
 $\langle \Psi | \Psi^\prime \rangle = \det \mathbf{\Psi}^\dagger \mathbf{\Psi}^\prime$; computing the matrix $\mathbf{\Psi}^\dagger \mathbf{\Psi}^\prime$ is of the order 
 of $N_p^2 N_s$ operations, while  the determinant can be computed using LU decomposition and takes on the order of $N_p^3$ operations. As long as the two Slater determinants are not orthogonal to each other, 
 the one-body density matrix is $\rho_{ab} = \langle \Psi | \hat{c}^\dagger_a \hat{c}_b | \Psi \rangle = ( \mathbf{\Psi^\prime} (\mathbf{\Psi}^\dagger \mathbf{\Psi}^\prime)^{-1}
 \Psi^\prime)_{ba}$, and the (normalized) matrix element of a two-body operator $\hat{V}=\sum_{abcd} V_{abcd} \hat{c}^\dagger_a \hat{c}^\dagger_b \hat{c}_d \hat{c}_c $ 
 is $\langle \Psi | \hat{V}| \Psi \rangle/ \langle \Psi | \Psi^\prime \rangle = \sum_{abcd} V_{abcd} ( \rho_{ac} \rho_{bd} - \rho_{ad}\rho_{bc} )$. 
This last sum, which goes roughly like $N_s^4$ (though some matrix elements are zero by selection rules), and  because $N_s > N_p$ it is evalution of the Hamiltonian that is 
  computationally burdensome. 
For more detailed exposition on this, see \cite{PhysRevC.66.034301,PhysRevC.48.1518} and references therein. In other frameworks, i.e., coordinate space mean-field calculations, the 
analysis may be different.

As in our implementation, the norm matrix elements are far cheaper than the Hamiltonian,
 we devised a `need to know' methodology:  (1) using a large $J_\mathrm{max}$, compute the norm matrix elements and the $f_J$, confirming that the 
 sum rule (\ref{sumrule}) is satisfied; (2) using some cutoff $f_\mathrm{min}$, selected the occupied values of $J$ such that $f_J > f_\mathrm{min}$, typically 
 around $10^{-3}$; (3) solve 
 (\ref{linrelnorm}) and (\ref{linrelham}) but using only the occupied values of $J$ in the sum.  This means fewer terms in the expansion and thus 
 a corresponding smaller number of Euler angles $\Omega_i$ at which to evaluate the computationally expensive left-hand side of (\ref{linrelham}) in particular.
 
We can sketch out the comparative computational burden.  For most of our cases, we found quadrature meshes 
 between
 $25^3 = 15,625$ Euler angles and  $32^3=32,768$ angles were sufficient.   Suppose we have a $J_\mathrm{max} = 12$. 
 In the simplest possible mesh for linear algebra projection (LAP),  $J$ runs from 0 to 12, and f $M, K$ run between -12 and 12; thus the number of evaluations are 
 $13 \times 25 \times 25 = 8125$ Euler angles, or lbetween half and a quarter as many evaluations required as for quadrature. In fact, this is overcounting: the minimal number of 
 evaluations should be, for each occupied value of $J$, a sum over all combinations of allowed $M$ and $K$ or $2J+1$ for each, that is, $\sum_{J=0}^{J_\mathrm{max}} (2J+1)^2 \approx (4/3) J_\mathrm{max}^3$ which is a factor of 3 smaller still.  While we have not yet implemented 
 such a minimal mesh for LAP, we did implement a `need to know' mesh. For example, if one has only even $J$ values, then the number of evaluations is half as much. This involves inverting $\Delta_{KM}^{J^\prime J}$ only for select values of $J^\prime, J$. We gave such an example for $^{48}$Cr in section \ref{examples}.   
 
 We found, however, the invertibility of the  matrix $\Delta_{KM}^{J^\prime J}$ as defined in Eq.~(\ref{DeltaJJdefn}) is surprisingly sensitive to both the choice of $J$s and to the 
 angles $\beta_i$.  We succeeded in the case of $^{48}$Cr, by taking only the even values of $J, J^\prime$ and skipping every other value of 
 $\beta_i$ in the mesh (\ref{betamesh}), but in other cases we were not successful.  Fortunately the invertibility is easy to know, as it 
 depends upon the eigenvalues of the symmetric matrix $\Delta_{KM}^{J^\prime J}$. We found as long as the eigenvalues were $> 10^{-4}$ we 
 got good results.  While further investigation is needed, this approach looks promising.

\section{Conclusions and future work}

We have proposed and demonstrated a new method of projecting angular momentum using linear algebra.  Our initial implementation demonstrates the method works and, for moderately high angular momentum $J$, is computationally competitive, in many cases requiring significantly fewer  evaluations than 
standard quadrature methods.   While for our demonstrations we mostly used the most straightforward separable inversion, 
we demonstrated a `need-to-know' inversion, where one computed the fractional occupations $f_J$ via the norm matrix and then used a reduced 
sampling for extracting the computationally expensive Hamiltonian matrix.  Because the inversion becomes sensitive to the choice of angles, further investigation into 
these improved inversions is suggested. 
We also leave to future work 
application to systems beyond shell-model spaces (i.e., coordinate-space mean-field wave functions),  to cases with multiple initial states, e.g., generator-coordinator and 
relative methods, to transitions, 
and  other finite quantum numbers such as isospin and 
particles number \cite{PhysRevC.92.064310}.

This material is based upon work supported by the U.S. Department of Energy, Office of Science, Office of Nuclear Physics, 
under Award Number  DE-FG02-96ER40985, as well as internal funding from San Diego State University to support one of us (O'Mara) for summer research. 
 The implementation of angular momentum projection via quadrature was modified from a prior, unpublished version by J.~T.~Staker.  CWJ would also like 
 to thank Changfeng Jiao for inspiring discussions.

\bibliographystyle{apsrev4-1}

\bibliography{johnsonmaster}

\end{document}